\documentstyle[prb,aps,twocolumn,amstex]{revtex}

\begin{document}

\draft
\title{Interface magnetic anisotropy in cobalt clusters embedded in a 
platinum 
or niobium matrix.}
\author{M. Jamet$^{1}$, M. N\'egrier$^{1}$, V. Dupuis$^{1}$, J. 
Tuaillon-Combes$^{1}$, P. M\'elinon$^{1}$, A. P\'erez$^{1}$, W. 
Wernsdorfer$^{2}$, B. Barbara$^{2}$, J. Vogel$^{2}$ and B. Baguenard$^{3}$}
\address{$^1$ D\'epartement de Physique 
des Mat\'eriaux, Universit\'e Claude Bernard-Lyon 1 et CNRS, 69622 
Villeurbanne, FRANCE. \\
 $^2$ Laboratoire Louis N\'eel, CNRS, 38042 
Grenoble, FRANCE.  \\
$^3$ Laboratoire de Spectrom\'etrie ionique et mol\'eculaire, Universit\'e 
Claude Bernard Lyon 1 et CNRS, 69622 Villeurbanne, FRANCE.}
\date{\today}
\maketitle

\begin{abstract}
A low concentration of cobalt clusters with a fcc structure and containing 
almost one thousand atoms are embedded in two different metallic matrices: 
platinum and niobium. Samples have been prepared using a co-deposition 
technique. Cobalt clusters preformed in the gas phase and matrix atoms are 
simultaneously deposited on a silicon substrate under Ultra High Vacuum 
(UHV) conditions. This original technique allows to prepare nanostructured 
systems from miscible 
elements such as Co/Pt and Co/Nb in which clusters keep a pure cobalt core 
surrounded with an alloyed interface. Magnetic measurements performed 
using a Vibrating Sample Magnetometer (VSM) reveal 
large differences in the magnetic properties of cobalt clusters in Pt 
and Nb pointing out the key role of cluster/matrix interfaces. 
\end{abstract}
\bigskip
\narrowtext
\section {Introduction}

	Magnetic nanostructures are subjects of growing interest on account of 
their potential applications in the fields of high density magnetic 
recording media and spin electronics. Indeed, using magnetic clusters 
(containing from a few hundreds to a few thousand atoms) as memory bits 
should highly 
increase the storage density. However, clusters with a 
large magnetic anisotropy have to be used in this case to overcome the 
superparamagnetic limit\cite{Dorm97}. Cobalt clusters embedded in a 
platinum matrix could be good 
candidates because Co/Pt multilayers display very large perpendicular 
magnetic 
anisotropy (PMA) \cite{Naka98}. Despite the nearly spherical shape of 
 clusters, a large remaining interface magnetic anisotropy is expected. 
Therefore, this system would allow to increase the particle blocking 
temperature up to temperatures compatible with magnetic recording 
applications. In the same time, we are investigating the magnetic 
properties of individual cobalt 
clusters using a new microSQUID technique\cite{encours00} to rule out any 
statistical treatment of the experimental data (taking into account size, 
shape or defects distributions). Note that Co/Nb multilayers 
are also used in superconducting spin valve devices\cite{Dieny}. For these 
reasons, a detailed study of Co clusters embedded 
in a superconducting niobium matrix is reported in a previous 
paper\cite{Jame00}. Though 
cobalt clusters have similar structures in Co/Pt and Co/Nb systems (a 
pure  
cobalt core and an alloyed interface), we observe that their magnetic 
properties are drastically different. In the present paper, we first 
report 
a detailed magnetic study of the Co/Pt system showing an anomalous 
dependence of the cluster magnetization with temperature and a large 
interface anisotropy. Then, we compare these results with previous 
ones obtained with the Co/Nb system. 

\section {Sample preparation and structural characterization}

	Cobalt clusters are prepared using a laser vaporization source 
improved according to Milani-de Heer design\cite{Mila90}. The Ti:sapphire 
vaporization 
laser used provides output energies up to 300 mJ at 790 nm, for 
a pulse duration of 3 $\mu s$ and a 10 Hz repetition rate\cite{Pell94}. 
This source produces an intense supersonic cluster beam allowing us to 
grow films in the Low Energy Cluster Beam Deposition 
(LECBD) regime. In this case, clusters do not fragment upon impact on the 
substrate or in the matrix\cite{Pere97}. We can prepare nanostructured 
thin films by 
random stacking of incident free clusters on different substrates or 
films of cobalt clusters embedded in various matrices (here Pt and Nb) 
thanks to the co-deposition technique. This last one consists in two 
independent beams reaching at the same time a silicon (100) substrate 
tilted at 45$^{\circ}$ at room temperature: the preformed neutral cluster 
beam and 
the atomic beam used for the matrix. Depositions are performed in a UHV 
chamber 
(p=5$\times$10$^{-10}$ Torr) to limit cluster and matrix contamination. 
The 
matrix is evaporated using a UHV electron gun evaporator mounted in
the deposition chamber. By controlling both evaporation rates with quartz 
balance monitors, we can continuously adjust the cluster concentration in 
the matrix. Moreover, few neutral cobalt clusters (thickness e $<$ 1 
monolayer) are 
deposited on a carbon coated copper grid and subsequently protected by a 
thin amorphous carbon layer on top (10 nm) 
to perform ex-situ High Resolution Transmission Electron Microscopy 
(HRTEM) 
observations. Nearly spherical clusters with a 
fcc structure and a rather sharp size distribution (mean diameter 
$D_{m}\approx 3.0$ nm, dispersion $\sigma=0.2-0.3$) are observed. 
Actually, in order to 
minimize their surface energy, clusters mainly have a truncated octahedron 
shape\cite{Butt83}. A 20 nm-thick film of randomly stacked cobalt 
clusters on a silicon substrate is prepared to perform Grazing Incidence 
Small Angle X-ray Diffraction (GISAXD) measurements at LURE (Laboratoire 
pour l'Utilisation du Rayonnement Electromagn\'etique, Orsay-FRANCE). The 
diffraction spectrum reported in Fig. 1 clearly confirms that cobalt 
clusters exhibit a fcc structure with roughly the same mean diameter 
(D$_{m}$ $\approx$ 
3.9 nm) as the one derived from TEM observations.\\
In order to investigate the crystallographic structure of cobalt clusters 
embedded in Pt and Nb matrices, we prepared 500 nm-thick Pt and Nb films 
containing a 15 $\%$ volumic concentration of clusters. We recorded and 
compared the 
GISAXD signals for two different photon energies: h$\nu=$7.7 keV (cobalt 
K-edge) and h$\nu=$7 keV 
in Co/Nb. 
For h$\nu=$7 keV, we still observe the cobalt fcc (111) peak as in the 
pure cobalt cluster film (Fig. 
1). Note that we previously obtained the same result for cobalt clusters 
embedded in a SiO$_{x}$ matrix\cite{Tuai95}. Unfortunately, in the 
particular case of Co/Pt, the 
X-ray contrast between Co and Pt is so large that we cannot observe the 
cobalt signal. As a consequence, we will assume that cobalt clusters keep 
their fcc structure in Co/Pt as in Co/Nb. Furthermore, previous X-ray 
absorption measurements (Extended X-ray Absorption Fine Structure: EXAFS) 
were performed at LURE on Co/Pt\cite{Negr00} (resp. Co/Nb\cite{Jame00}) 
systems. They revealed an alloying effect at the cluster surface in both 
Co/Pt and Co/Nb miscible systems. For magnetic measurements using a VSM 
apparatus, we prepared a 500 nm-thick 
Pt film containing a 4 $\%$ cobalt volume
concentration (Co/Pt sample) and a 500 nm-thick Nb film containing a 2 
$\%$ cobalt volume concentration (Co/Nb sample). Very low concentrations 
are chosen to rule out any magnetic 
coupling between particles in the sample.

\section {Magnetic study of the Co/Pt system}

	In the following, we note M$_{s}$(T) the particle magnetization which is 
temperature dependent. 
This parameter is important since it gives information on the magnetic 
state of the particles. The magnetization is proportionnal to the 
saturation 
moment of the sample labeled m$_{sat}$(T) through the relation:

\begin{equation}
m_{sat}(T)=M_{s}(T)\int_{0}^{\infty}{N(\pi D^{3}/6)f(D)\, dD} \propto 
M_{s}(T)
\end{equation}

Particles are assumed to be nearly spherical and D is the diameter, N is 
the total number of particles in the sample and f(D) 
is the log-normal magnetic size distribution with D$_{m}$ the mean 
diameter and $\sigma$ the dispersion:

\begin{equation}
f(D)=\frac{1}{D\sqrt{2\pi\sigma^{2}}}exp\Biggl(-\biggl(ln\Bigl(\frac{D}{D_{m}}\Bigr)\biggr)^{2}\frac{1}{2\sigma^{2}}\Biggr)
\end{equation}

Considering the 
narrow size distribution of clusters as deduced from TEM observations, we 
assume that the 
magnetization does not depend on the particle size in Eq. (1) 
(ref. 12). Moreover, in the following, we consider 
noninteracting particles {\textit 
i.e.} only the applied magnetic field ({\textbf H$_{app}$}) has to be 
taken into account. 
Furthermore for cobalt, the exchange length is 7 nm which is larger than 
the mean 
particle size of 3 nm. Thus, a single domain cluster can be seen as a 
macrospin with uniform rotation of its magnetization. Finally, we will 
assume a uniaxial magnetic anisotropy within the particles\cite{comm}. If 
the easy magnetization directions are randomly distributed and the applied 
magnetic field is larger than the coercive field H$_{c}$(T), the magnetic 
moment of the sample can be written as\cite{Dorm97}:

\begin{align}
m(H&_{app},T)=m_{sat}(T)\ \Biggl<\int_{0}^{\pi}d\psi \ \frac{sin\psi}{2}\\ 
\nonumber &\frac{\int_{0}^{\pi}d\theta \ sin\theta \int_{0}^{2\pi}d\phi \ cos\alpha \ 
exp\ (\eta\ cos^{2}\theta +\xi\ cos\alpha)}{\int_{0}^{\pi}d\theta \ 
sin\theta \int_{0}^{2\pi}d\phi \ exp\ (\eta\ cos^{2}\theta +\xi\ 
cos\alpha)}\Biggr>
\end{align}

Here, we use the spherical coordinates. The two angles ($\theta$,$\phi$) 
give the direction of the particle magnetization. The easy magnetization 
direction is fixed along the $z$ axis while the magnetic field orientation 
given by the angle $\psi$ between \textbf{H}$_{app}$ and the easy 
direction is continuously varied from $\psi=0$ to $\psi=\pi$. 
$cos\alpha =sin\psi\ sin\theta\ sin\phi +cos\psi\ cos\theta$, $\eta=\Delta 
E/k_{B}T$ where $\Delta E$ is the energy barrier to cross in order to 
reverse the particle magnetization, and $\xi =\mu_{0}H_{app}(\pi 
D^{3}/6)M_{s}(T)/k_{B}T$. The final expression is then averaged over the 
magnetic size distribution $f(D)$ according to the formula:

\begin{equation}
<\Gamma >\ =\ \frac{\int_{0}^{\infty}D^{3}\Gamma(D) f(D)\, 
dD}{\int_{0}^{\infty}D^{3}f(D)\, dD}
\end{equation}

where $\Gamma(D)$ is a function of the particle diameter. Eq. (3) 
simplifies in the case $\xi>>1$ (high field, low 
temperature)\cite{Dorm97}:

\begin{equation}
m(H_{app},T)\approx m_{sat}(T)
\biggl(1-\biggl<\frac{1}{\xi}\biggr>-\frac{4}{15}\biggl<\frac{\eta^{2}}{\xi^{2}}\biggr>+...\biggr)
\end{equation}

In a first approximation, we use this expansion limited to the first order 
as a saturation approach law to determine m$_{sat}$(T):

\begin{align}
m(H_{app},T)&\approx m_{sat}(T)\ 
\biggl(1-\frac{a}{\mu_{0}H_{app}}\biggr),\\ \nonumber 
a&=\frac{k_{B}T\ exp\ (-9\sigma^{2}/2)}{(\pi D_{m}^{3}/6)\ M_{s}(T)}
\end{align}

Since we assume that the magnetization is independent on the 
applied magnetic field, we certainly give an upper limit for m$_{sat}$(T). 
From the magnetization curves reported in Fig. 2, we clearly show that the 
saturation moment is temperature 
dependent. Thus the magnetization also depends on the 
temperature and its evolution vs. T is shown in Fig. 3. At very low 
temperature, the thermal fluctuations of the magnetization are negligible 
and we obtain a relevant information on the magnetic state of the cluster: 
$M_{s}(0K)\approx M_{s}(1.5K)=1600\pm200\ kA.m^{-1}$. This value is larger 
than the cobalt bulk magnetization ($M_{s}^{bulk}=1430\ kA.m^{-1}$). \\
 When the coercive field H$_{c}$(T) is zero, Eq. (3) also simplifies in 
the case $\xi <<1$ (low field, high temperature), and can be written to 
the first order\cite{Mull73}:

\begin{align}
m(H&_{app},T)\approx m_{sat}(T)\ \biggl<\frac{\xi}{3}\biggr>\\ \nonumber
          &= m_{sat}(T)\ \frac{\mu_{0}H_{app}(\pi D_{m}^{3}/6)
M_{s}(T)}{3k_{B}T}\ exp(13.5\ \sigma^{2})
\end{align}

and using Eq. (1), we find:

\begin{equation}
m(H_{app},T)=\frac{N(\pi D_{m}^{3}/6)^{2}exp(18\sigma^{2})}{3k_{B}}
\frac{\mu_{0}H_{app}M_{s}^{2}(T)}{T}
\end{equation}

This corresponds exactly to the magnetic moment we measure in the Zero 
Field Cooled 
(ZFC) protocole when all the particles are superparamagnetic (the remanent 
moment of the sample being equal to zero). Thus, we deduce: 
$M_{s}(T)\propto\sqrt{T\ m_{ZFC}(H_{app},T)/\mu_{0}H_{app}}$ (ref. 
15). In Fig. 3, we plot $\sqrt{T\ 
m_{ZFC}(H_{app},T)/\mu_{0}H_{app}}$ for 3 different applied magnetic 
fields. The resulting curves superimpose with the one obtained from 
m$_{sat}$(T) for T$>$150 K in the superparamagnetic regime. In this field 
range, we notice that the magnetization does not depend on 
$\mu_{0}H_{app}$.
In the following, the temperature dependence of the particle magnetization 
is systematically taken into account. We first estimate the magnetic size 
distribution from the highest temperature magnetization curves. At T=300 K 
and T=250 K, anisotropy terms can be neglected and a simple Langevin 
function $L(\xi)$ allows us to fit m(H$_{app}$,T) (ref. 5) (Fig. 4):

\begin{equation}
\frac{m(H_{app},T)}{m_{sat}(T)}=\frac{\int_{0}^{\infty}D^{3}L(\xi)f(D)\, 
dD}{\int_{0}^{\infty}D^{3}f(D)\, dD}
\end{equation}

We deduce: $D_{m}=2.7\pm 0.1\ nm$ and $\sigma=0.35\pm 0.05$ which roughly 
corresponds to the size distribution obtained from TEM observations. 
Decreasing the temperature, anisotropy terms are no more negligible and 
have to be considered to fit the experimental data. To estimate them, we 
perform a detailed analysis of both the remanent moment vs. temperature 
and the ZFC magnetization curves using the magnetic size distribution 
previously found. \\
From hysteresis loops at low temperature, we deduce the remanent moment 
m$_{r}$(T) of the sample. We do not observe a narrowing of the loops at 
the vicinity of $\mu_{0}H_{app}=0\ T$ below 10 K so that we can assume 
that almost all the cobalt particles are magnetically blocked below this 
temperature. 
Moreover, we notice that $m_{r}(T)\approx m_{sat}(T)/2$ which implies a 
uniaxial magnetic anisotropy within the particles\cite{Pfei90}. In Fig. 
5(a), 
we plot $m_{r}(T)/m_{r}(1.5K)$ and fit this curve using the following 
expression\cite{Jame00}:

\begin{equation}
\frac{m_{r}(T)}{m_{r}(1.5K)}=\frac{\int_{D_{B}(T)}^{\infty}D^{3}f(D)\, 
dD}{\int_{D_{B}(1.5K)}^{\infty}D^{3}f(D)\, dD}
\end{equation}

where D$_{B}$(T) is the blocking diameter of the particles at temperature 
T and $\mu_{0}H_{app}=0\ T$. One finds D$_{B}$(T) when the relaxation time 
of the particle is equal to the measuring time: $\tau=\tau_{0}\ exp\ 
(\Delta E/k_{B}T)=\tau_{mes} \Leftrightarrow \Delta E=k_{B}T\ 
ln(\tau_{mes}/\tau_{0})$. In our case, $\tau_{mes}=10\ s$ and the attempt 
frequency $\tau_{0}^{-1}$ is typically 10$^{9}$-10$^{12}$ Hz. To fit the 
ratio $m_{r}(T)/m_{r}(1.5K)$, we take $\Delta E=K\ D^{\alpha}$ where $K$ 
and $\alpha$ are free parameters. The best fitting values are given in 
Table I, they correspond to an interface anisotropy with $\alpha\approx 
2$. We also fit the ZFC magnetization curves for different applied 
magnetic 
fields using (Fig. 5(b)):

\begin{equation}
\frac{m_{ZFC}(H_{app},T)}{m_{sat}(T)}=\frac{\int_{0}^{D_{B}(H_{app},T)}D^{3}(\xi 
/3)f(D)\, dD}{\int_{0}^{\infty}D^{3}f(D)\, dD}
\end{equation}

Here, we neglect the blocked particle susceptibility. The blocking 
diameter D$_{B}$(H$_{app}$,T) now depends on the applied magnetic field, 
and the anisotropy energy barrier is written: $\Delta E=K(H_{app})\ 
D^{\alpha}$ where the anisotropy constant $K(H_{app})$ which may depend on 
the applied magnetic field and the exponent $\alpha$ are free parameters. 
The results are given in Table I for six different magnetic fields. They 
show that the anisotropy constant is actually independent on the magnetic 
field and confirm an interface anisotropy within the particles: 
$\alpha\approx 2$. \\
At intermediate temperatures: T$=$200 K and T$=$150 K, we cannot fit the 
magnetization curves using a simple Langevin function as shown in Fig. 6. 
Indeed to achieve this, one has to use Eq. (3) to take the anisotropy term 
into account\cite{Hans93} (i.e. $\Delta E=K\ D^{\alpha}$). Thus we assume 
an interface anisotropy as previously suggested and solve numerically Eq. 
(3) in order to deduce the anisotropy constant. Fig. 6 
shows theoretical and experimental curves, and the resulting anisotropy 
constants are given in Table I. From all these results we can assert that 
interface is responsible for the large uniaxial magnetic anisotropy in 
cobalt clusters embedded in a platinum matrix. Note that it was not 
necessary to take the temperature dependence of the anisotropy constant 
into account to fit the experimental data. It is now interesting to compare
 this system with Co/Nb.

\section {Comparison with the Co/Nb system}

	Here, we summarize briefly the main results we obtained on the 
Co/Nb sample in a previous paper\cite{Jame00}. Magnetization 
measurements are performed at temperatures higher than 8 K because of the 
superconducting behavior of niobium films for T$<$7 K. As in the case of 
Co/Pt, we use Eq. (6) 
as a saturation approach law to determine m$_{sat}$(T) and subsequently 
Eq. (1) to deduce the cobalt magnetization M$_{s}$(T) in the sample (Fig. 
7). In the same figure, we 
plot $\sqrt{T\ m_{ZFC}(H_{app},T)/\mu_{0}H_{app}}$ for 3 different applied 
magnetic fields and T$<$150 K (above this temperature the magnetic signal 
reaches the magnetometer resolution). We find the 
magnetization: $M_{s}=500\pm 50\ kA.m^{-1}$ which shows nearly no 
dependence on the temperature. The same magnetic measurements performed 
with 
a second sample containing a 3 $\%$ volume concentration of cobalt 
clusters give: 
$M_{s}=510\pm 60\ kA.m^{-1}$. Finally, complementary X-ray Magnetic 
Circular Dichroism 
(XMCD) was performed at the ESRF (European Synchrotron Radiation Facility, 
Grenoble-FRANCE). The absorption signal 
was recorded under a 3 Tesla magnetic field at T$=$5 K, and we measured 
the total electron yield in the photon energy range which spans the L2 and 
L3 absorption lines of cobalt. Using the sum rules\cite{Thol92}, we can 
estimate the mean atomic magnetic moment of cobalt atoms: $\mu_{at}=0.5\pm 
0.05\ \mu_{B}$ which corresponds to $M_{s}=430\pm 50\ kA.m^{-1}$. This 
value is in good agreement with the previous ones. Moreover, the orbital 
to spin moment ratio is enhanced compared with the bulk value\cite{XMCD} 
as expected in small particles: $m_{L}/m_{S}\approx 0.2$. However, one has 
to notice that this technique probes the first 3 nm of the sample where 
slight 
oxidation might be possible. \\
We note $D_{m}^{mag}$ and $\sigma^{mag}$ (resp. $D_{m}^{TEM}$ and 
$\sigma^{TEM}$) the mean cluster diameter and the dispersion of the 
magnetic (resp. TEM) size distribution. If we assume that: 
$\sigma^{mag}=\sigma^{TEM}$, one can write: 
$M_{s}=M_{s}^{bulk}(D_{m}^{mag}/D_{m}^{TEM})^{3}$. By taking $M_{s}=500\pm 
50\ kA.m^{-1}$ and $D_{m}^{TEM}=3.0\pm 0.1\ nm$, we find 
$D_{m}^{mag}=2.1\pm 0.2\ nm$. This is in rather good agreement with the 
magnetic size: $D_{m}=2.1\pm 0.1\ nm$, $\sigma=0.35\pm 0.05$ which fits 
really well the magnetization curves at T$=$300 K and T$=$200 K using the 
bulk magnetization (Fig. 8). We first notice a large difference between 
the magnetic mean diameter in Co/Pt (2.7 nm) and the one in Co/Nb (2.1 
nm). In the case of Co/Nb, we conclude that at least two cobalt atomic 
monolayers are magnetically dead at the cluster surface (Fig. 9(a) and 
9(b)) while 
in Co/Pt, the alloying effect is limited since we roughly find the same 
diameter as the one derived from TEM 
observations. In ref. 5 concerned with Co/Nb, we only considered a volume 
anisotropy K$_{V}$ within cobalt 
clusters and found: $K_{V}=2.0\pm 0.3.10^{5}\ J/m^{3}$. In the present 
work, we assume that cobalt clusters exhibit an interface anisotropy in 
Co/Nb to allow a comparison with Co/Pt. In this case, the anisotropy 
energy barrier is 
written: $\Delta E=K_{S}D^{2}$. We use Eq. (10) to fit the evolution of 
the remanent moment vs. temperature (Fig. 10(a)) and Eq. (11) to fit the 
ZFC magnetization curves (Fig. 10(b)). In both cases, we find: 
$K_{S}=0.05\pm 
0.008\ mJ/m^{2}$ which is almost one order of magnitude smaller than in 
Co/Pt. 

\section {Discussion and conclusion}

Cobalt-platinum and cobalt-niobium elements are miscible which may promote 
interdiffusions at 
the cluster/matrix interfaces. Indeed, previous structural 
studies\cite{Negr00,Jame00} showed that one (resp. two) cobalt atomic 
layer 
diffuses inside the Pt (resp. Nb) matrix. In this discussion we show that 
a simple core-shell model with a pure cobalt core surrounded with a 
disordered Co$_{x}$Pt$_{1-x}$ or Co$_{x}$Nb$_{1-x}$ alloyed shell can 
originally account for the magnetic properties of cobalt clusters embedded 
in both Pt and Nb matrices. A comparable model has already been suggested 
by 
Canedy \textit{et al.}\cite{Cane00} in Co/Pt superlattices. We can write 
the cluster magnetization as:

\begin{equation}
M_{s}(T)=xM_{s}^{core}(T)+(1-x)M_{s}^{shell}(T)
\end{equation}

where $x$ is the fraction of cobalt atoms in the core with 
$M_{s}^{core}(T)=M_{s}^{bulk}(T)$ ($M_{s}(0K)=1430\ kA.m^{-1}$), and 
$(1-x)$ the fraction of cobalt atoms in the alloyed interface shell. For 
the following calculations we consider the cluster model reported in Fig. 
9(a) and 9(b) which 
contains 1289 atoms. For Co/Pt, one atomic layer is expected to diffuse 
inside the matrix giving $x=0.63$. Thus one finds: $M_{s}(0K)=0.63\times 
1430+0.37\times M_{s}^{shell}(0K)=1600\ kA.m^{-1}$ which provides: 
$M_{s}^{shell}(0K)\approx 1900\ kA.m^{-1}$. This magnetization enhancement 
of cobalt is much smaller than that in CoPt alloys (3325 kA.m$^{-1}$) 
(ref. 20). This may be due to dimensional effects as mentioned by Canedy 
\textit{et 
al.}\cite{Cane00}. However, the temperature dependence of the 
magnetization indicates that the Curie temperature of the interface 
($T_{C}$) is quite small (the Curie temperature in the core is assumed to 
be much higher than 300 K). Besides, above this temperature the sample 
magnetization reaches: $M_{s}(T)=xM_{s}^{bulk}(T)\approx 0.63\times 
1430=901\ kA.m^{-1}$. Experimentally, that gives: $200<T_{C}<250\ K$ (Fig. 
3). The low Curie temperature of CoPt disordered alloys has already been 
pointed out by Weller \textit{et al.}\cite{Well92} or 
Devolder\cite{Devo00}.\\
We further annealed this sample during 10 minutes 
at T$=$450$^{\circ}$C under a vacuum of 10$^{-7}$ Torr. In Fig. 3, we plot 
the sample 
magnetization vs. T using the saturation approach law given by Eq. (6). At 
2 K, the large magnetization $M_{s}=2720\ kA.m^{-1}$ now approaches the 
CoPt alloy value. It implies that almost all the cluster atoms have 
diffused inside the platinum matrix to form small alloyed "clusters". The 
low Curie temperature: $T_{C}\approx 150\ K$ confirms that there is no 
more pure cobalt core in the sample ($x=0$). Moreover, magnetization 
curves measured down to 2 K show no remanent moment so that the alloyed 
"cluster" anisotropy is negligible as expected for disordered 
CoPt alloys\cite{Devo00}.\\
For Co/Nb, two atomic layers are expected to 
diffuse inside the matrix thus: $x=0.36$. If we assume that the CoNb 
disordered alloy is magnetically dead\cite{Jame00}, one finds the sample 
magnetization: $M_{s}(T)=xM_{s}^{bulk}(T)\approx 0.36\times 1430=515\ 
kA.m^{-1}$, which is in good agreement with the experimental data (Fig. 
7). 
Finally, experimental magnetization values in Co/Pt and Co/Nb samples can 
be 
well interpreted on the basis of a simple core-shell model.\\
In Co/Pt, we unambiguously find the existence of an interface anisotropy 
and it is actually 
impossible to fit the experimental data by only considering a volume 
anisotropy within 
the clusters. Interface anisotropy originates from the combination of the 
large spin-orbit coupling in Pt with the natural anisotropy directions 
induced by the surface\cite{Naka98}. Further surface strains induced by 
the surrounding Pt atoms may contribute to this interface anisotropy. \\
Let us assume a volume anisotropy within the cobalt clusters. That leads 
to an anisotropy constant $K_{V}$ given by the relation: 
$K_{V}V_{m}=K_{S}S_{m}$ where $V_{m}$ and $S_{m}$ are the mean cluster 
volume and surface respectively. Using the mean diameter $D_{m}$ obtained 
in section III, one finds: $K_{V}=6K_{S}/D_{m}\approx 7.10^{5}\ J/m^{3}$. 
For an infinite cobalt cylinder, shape anisotropy is equal to 
$\mu_{0}M_{s}^{2}/4\approx 6.4.10^{5}\ J/m^{3}$ if we use the bulk 
magnetization\cite{Ahar96}. This is even smaller than the above K$_{V}$ 
value thus shape anisotropy cannot account for the experimental 
one. Furthermore, cubic magnetocrystalline anisotropy reported in 
ref. 24 ($1.2.10^{5}\ J/m^{3}$) or in ref. 25 ($2.7.10^{5}\ J/m^{3}$) are 
also too small to account for the experimental value. In conclusion the 
assumption of a volume anisotropy seems not physically obvious. \\
In order to compare our surface anisotropy with previous works, we can 
estimate the corresponding anisotropy energy per cobalt atom at the 
cluster surface. For this purpose, we use the cluster model in Fig. 9(c). 
Indeed, for a perfect truncated octahedron as the one given in Fig. 9(a) 
and 9(b), 
symmetries cancel surface anisotropy and we add one (111) facet to break 
this symmetry. We note $K^{at}$ the atomic anisotropy energy. Summing over 
all the cluster surface atoms, one finds the anisotropy energy in the 
whole particle: $E=-15K^{at}cos^{2}(\theta)$ where $\theta$ is the angle 
between the magnetization and the [111] direction. Dividing by the cluster 
surface, we can deduce $K^{at}$ from the experimental anisotropy: 
$K^{at}\approx 3\ meV/at$. This is one order of magnitude larger than the 
experimental values given in ref. 2, 26, 27 ranging between 0.1 and 0.3 
meV/at for 1-2 cobalt monolayers in Co/Pt multilayers. However, 
comparisons with Co/Pt multilayers are quite difficult because this is 
more a two-dimensional problem whereas clusters involve a 
three-dimensional treatment.
Concerning Co/Nb, we believe that interface anisotropy is so small that it 
becomes possible to fit the magnetization curves using a volume 
anisotropy. But recent works performed on single cobalt 
clusters\cite{encours00} embedded in a niobium matrix show that interface 
still rules magnetic anisotropy in this system.\\
Other experimental issues to increase magnetic anisotropy in nanosized 
particles and consequently the corresponding blocking temperature could be 
tested in the near future: using another matrix element to increase 
spin-orbit coupling at the cluster surface, preparing slightly elongated 
particles since interface anisotropy is proportional to the deformation... 

\section{Aknowledgements}

The authors would like to thank Professor C. BINNS from the Department of 
Physics and Astronomy-University of Leceister-UK, 
for his help during the XMCD measurements on the ID12B line 
at the ESRF.

\begin{figure}

\caption{(a) Diffraction pattern (GISAXD) obtained on a 20 nm-thick film 
of 
randomly stacked cobalt clusters (crosses). Supported clusters clearly 
exhibit a fcc 
structure. (b) XRD spectra obtained on a 500 
nm-thick niobium film containing a 15 $\%$ volume concentration of cobalt 
clusters for two different photon energies: h$\nu=$7.7 keV (cobalt K-edge) 
(solid line) and h$\nu=$7 keV (empty circles). Even embedded in a niobium 
matrix cobalt clusters still exhibit a fcc structure.}

\label{fig 1}
\end{figure}

\begin{figure}

\caption{Magnetization curves obtained for isolated cobalt clusters 
embedded in a platinum matrix (Co/Pt sample). Dots: experimental data, 
solid lines: 
approach to saturation simulated using Eq. (6). At T$=$1.5 K, in the 
ferromagnetic 
regime, the saturation moment is much higher than the one at T$=$300 K, in 
the superparamagnetic regime showing the large dependence of the 
magnetization on temperature.}

\label{fig 2}
\end{figure}

\begin{figure}

\caption{Experimental magnetization of the as-prepared and annealed Co/Pt
sample estimated using Eq. (6) as a saturation approach law (full 
circles). In the as-prepared sample, a core-shell cluster model with a 
pure cobalt core and an alloyed Co$_{x}$Pt$_{1-x}$ interface well accounts 
for the temperature dependence of the sample magnetization. We can deduce 
the interface Curie temperature: 200$<$T$_{C}<$250 K. In the annealed 
sample, the Curie temperature is approximately 150 K and M$_{s}$(0K) is 
much larger than in the as-prepared sample. We also plot $\sqrt{T\ 
m_{ZFC}(H_{app},T)/\mu_{0}H_{app}}$ which is proportional to M$_{s}$ in 
the superparamagnetic regime for three different applied magnetic fields: 
$\mu_{0}$H$_{app}=$10 mT; 12.5 mT and 15 mT (full squares). The three 
curves superimpose with the magnetization curve obtained from the 
saturation approach for T$>$150 K.}

\label{fig 3}
\end{figure}

\begin{figure}

\caption{Experimental magnetization curves in the superparamagnetic regime 
(T$=$250 K: full squares, T$=$300 K: full circles) for the Co/Pt sample. 
These curves are 
easily fitted using a simple Langevin function (solid lines) which 
gives the magnetic size distribution of cobalt clusters: 
D$_{m}$=2.7$\pm$0.1 nm and 
$\sigma$=0.35$\pm$0.05.}

\label{fig 4}
\end{figure}

\begin{figure}

\caption{(a) Experimental temperature dependence of the remanent moment 
(full circles) in 
the Co/Pt sample. The corresponding fitting curve (solid line) gives 
$\alpha$ and K in the anisotropy energy barrier (see Table I): $\Delta 
E=KD^{\alpha}$. (b) Experimental Zero Field Cooled (ZFC) magnetization 
curves given for 6 different applied magnetic fields. The fitting curves 
(solid lines) 
allow us to deduce $\alpha$ and K in the anisotropy energy barrier (see 
Table I). }

\label{fig 5}
\end{figure}

\begin{figure}

\caption{(a) Experimental magnetization curve at T$=$150 K (full squares) 
obtained for the Co/Pt sample. 
A simple Langevin function (dashed line) does not allow to fit the 
experimental data since the cluster anisotropy is no more negligible. 
Assuming an interface anisotropy, we can fit m/m$_{sat}$ (solid line) 
and deduce the anisotropy constant K (see Table I). (b) Experimental 
magnetization curve at T$=$200 K (full squares). A simple Langevin 
function (dashed line) does not allow to fit the experimental data since 
the cluster anisotropy is no more negligible. Assuming an interface 
anisotropy, we can fit m/m$_{sat}$ (solid line) and deduce the 
anisotropy constant K (see Table I).}

\label{fig 6}
\end{figure}

\begin{figure}

\caption{Experimental dependence of the magnetization vs. temperature 
using Eq. (6) 
as a saturation approach law (full circles) obtained for the Co/Nb sample. 
Note that the 
magnetization is nearly independent on temperature in this system. We also 
plot $\sqrt{T\ 
m_{ZFC}(H_{app},T)/\mu_{0}H_{app}}$ which is proportional to M$_{s}$(T) in 
the superparamagnetic regime for three different applied magnetic fields: 
$\mu_{0}$H$_{app}$=10 mT; 15 mT and 20 mT (full squares). The different 
curves superimpose for T$>$50 K in the superparamagnetic regime..}

\label{fig 7}
\end{figure}

\begin{figure}

\caption{Experimental magnetization curves in the superparamagnetic regime 
(T$=$200 K: full squares, T$=$300 K: full circles) for the Co/Nb sample. 
These curves are 
easily fitted using a simple Langevin function (solid lines) which 
gives the magnetic size distribution of the clusters: D$_{m}$=2.1$\pm$0.1 
nm and 
$\sigma$=0.35$\pm$0.05.}

\label{fig 8}
\end{figure}

\begin{figure}

\caption{(a) Model of cluster containing 1289 atoms with a truncated 
octahedron shape. (111) and (100) facets allow to minimize the cluster 
surface energy. (b) View along a [110] direction of the cobalt cluster. 
From the comparison of the magnetic size distribution with the TEM one, it 
seems that 
at least 2 atomic monolayers are magnetically dead at the cluster surface 
in Co/Nb. (c) Model of cluster containing 1337 atoms. Dark atoms belonging 
to the (111) facet are added to a perfect truncated octahedron basis of 
1289 atoms (light atoms) in order to 
break the cluster symmetry.}

\label{fig 9}
\end{figure}

\begin{figure}

\caption{(a) Experimental temperature dependence of the remanent moment in 
the Co/Nb sample (full circles). Assuming an interface anisotropy, the 
resulting fitting curve (solid line) gives the anisotropy constant 
K$_{S}$. (b) Experimental Zero Field Cooled (ZFC) magnetization curves 
obtained for 6 
different applied magnetic fields. The fitting curves (solid lines) 
allow us to deduce K$_{S}$.}

\label{fig 10}
\end{figure}

\begin{table}

\caption{$\alpha$ exponent and magnetic anisotropy constant K deduced from 
three different experimental measurements performed on the Co/Pt sample: 
remanent moment vs. temperature, Zero Field Cooled (ZFC) magnetization 
curves and magnetization curves at intermediate temperatures (see text). 
They give the anisotropy energy barrier: $\Delta E=KD^{\alpha}$ to cross 
in order to reverse the magnetization of a cobalt cluster with a diameter 
D.}

\label{TableI}
\end{table}


\begin{thebibliography}{26}
\bibitem{comm}
This assertion is justified in the next sections. A biaxial magnetic 
anisotropy should be a more general case. But the main difference only 
comes from the reversal path: the magnetization preferentially rotates 
perpendicularly to 
the hard magnetization axis in the case of a biaxial anisotropy.
\end{thebibliography}
\end{document}